\documentclass[aps,prl, footinbib,twocolumn,showpacs,amsmath,amssymb,superscriptaddress]{revtex4-2}
\usepackage{graphicx}
\usepackage{dcolumn}
\usepackage{bm}
\usepackage{hyperref}
\usepackage{multirow} 
\usepackage {color} 
\usepackage{amsmath, amssymb}
\usepackage{array}

\hypersetup{colorlinks=true, citecolor=blue, urlcolor=blue, linkcolor=blue}

\begin{document}

\title{Anomalous topological superradiant phases}
\author{You-Qi Lu}
\affiliation{Department of Physics, and Chongqing Key Laboratory for strongly coupled Physics,  Chongqing University, Chongqing 401330, China}
\author{Yu-Yu Zhang}
\email{yuyuzh@cqu.edu.cn}
\affiliation{Department of Physics, and Chongqing Key Laboratory for strongly coupled Physics,  Chongqing University, Chongqing 401330, China}
\author{Zi-Xiang Hu}
\email{zxhu@cqu.edu.cn}
\affiliation{Department of Physics, and Chongqing Key Laboratory for strongly coupled Physics,  Chongqing University, Chongqing 401330, China}

\date{\today}

\begin{abstract} 
We present a novel set of light-matter topology realized by implementing a finite-component quantum Rabi array with a photonic analog of the
Su-Schrieffer-Heeger (SSH) configuration. We demonstrate 
how complex light-matter couplings with species-dependent phases lead to the closure of superradiance-induced band gap in a
manner that differs from that in the SSH model. We uncover an topological superradiant phase transition from a normal phase to a topological superradiant electromagnet phase, which is characterized both by a local order parameter and a global topological invariant. Novel superradiance-enhanced edge states emerge with significantly amplified excitations superior to those in topological normal phase. Strikingly, tuning light-atom coupling induces novel topological superradiant electric and magnetic phases, exhibiting chiral edge-mode excitation at opposite boundaries. Our proposed setup offers a tunable platform for topological quantum optics, advancing applications in topological superradiant lasers.
\end{abstract}

\maketitle

\emph{Introduction --}The recent advances
in circuit QED have been widely studied
for simulating many-body quantum physics and gives rise to a wealth of strongly correlated light-matter phenomena~\cite{Natphys2009,RevModPhys.85.553,PhysRevLett.115.243604}. A hallmark effect of strongly coupled light–matter systems is the superradiant phase transition~\cite{PhysRev.93.99,PhysRevA.78.051801,PhysRevLett.124.073602}, which has been realized
in Bose-Einstein condensates~\cite{nature2010}. Recently, quantum Rabi model and two-site Jaynes–Cummings lattices in appropriate limits also exhibit a similar superradiant phase transition~\cite{PhysRevA.87.013826,PhysRevLett.115.180404,PhysRevLett.119.220601,PhysRevA.101.033827,NCcai2021,chen2021}. Such quantum phase transition in finite-component systems open a window for investigating related
exotic quantum phases, such as chiral and frustrated superradiant phases in extended quantum Rabi systems~\cite{PhysRevLett.127.063602,PhysRevLett.129.183602,PhysRevResearch.5.L042016,PRLZHANG}.

Introducing topology into light-matter interaction has advanced the emerging field of topological quantum optics~\cite{science2018,PhysRevLett.124.083603,RevModPhys.85.299}, where geometrical and topological properties are harnessed to control photons. Topological protection has been widely exploited in optical systems design to ensure robustness against local perturbations, as demonstrated by the realization of topological lasing~\cite{naturephoton,science2018E}, topological edge states of light against disorder~\cite{nature2013,naturephoton2013}, and robust photonic transport ~\cite{PhysRevLett.113.087403}. Extending this paradigm to strongly coupled light-matter systems offers a promising route toward superradiant photons, enabling a new generation
of superradiant lasing robust against local fluctuations~\cite{nature2012,sbbk-xdvs,science2018T}. Pioneering studies have demonstrated topological states with superradiance lattices~\cite{PhysRevLett.126.103601,PhysRevLett.134.193602}, and topological superradiant phases in a two-component Fermi gas~\cite{PhysRevLett.115.045303}. However, a key obstacle in strongly coupled light-matter systems is that particle-nonconserving interactions typically prevent band-gap closing, hindering the emergence of topology. The novel phenomenon of 
superradiant phase transitions with nontrivial topology remain largely unexplored.

We present anomalous topological superradiant phases in a finite-component quantum Rabi array by introducing electric and magnetic light-atom couplings on two sublattices, which circumvent the particle-nonconserving interactions. The superradiant transition becomes topologically modulated and the Zak phase are tunable by light-atom coupling, in contrast to the SSH model. We uncover topological superradiant phase transitions as a new paradigm, exhibiting dispersive edge modes and superradiance-enhanced excitations. Two distinct phases emerge, topological superradiant magnetic (TSM) and electric (TSE), featuring chiral edge states with excitations localized at opposite boundaries. This tunability establishes light-matter systems as a platform for engineering topological phases.
\begin{figure}
  \centering
  \includegraphics[width=\linewidth]{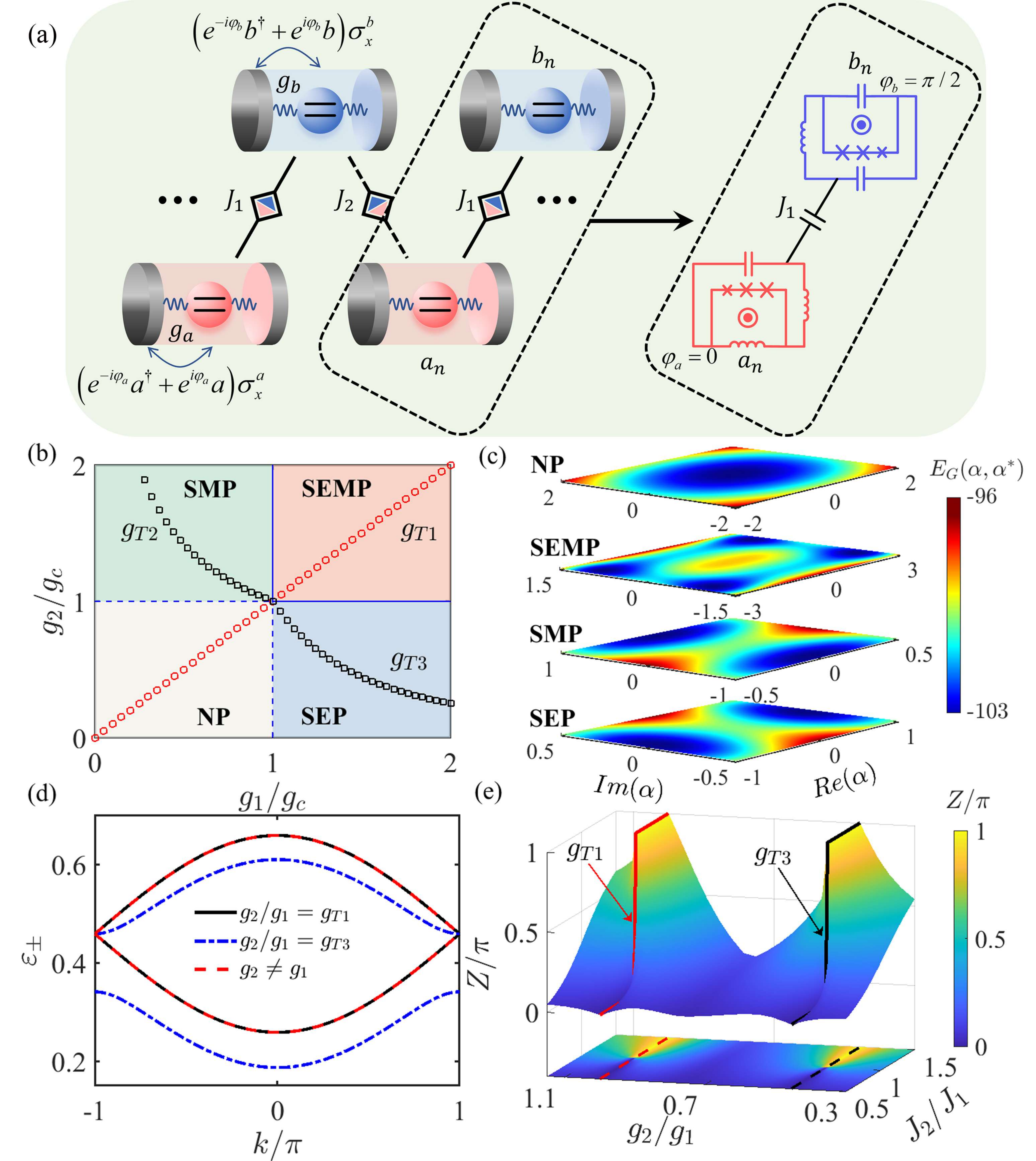}
  \caption{(a)Schematic of a quantum Rabi array with two sublattices A and B. Each Rabi unit is realized by a flux qubit coupled to an LC resonator via a capacitor or an inductor with geometric phases $\varphi_a=0$ or $\varphi_b=\pi/2$. (b) Phase diagram in the $g_2/g_c$-$g_1/g_c$ plane for $J_1=J_2=0.1$. Topological critical points are marked by $g_{\text{T1}}$ (red circles), $g_{\text{T2}}$ and $g_{\text{T3}}$ (black squares). Critical boundaries $g_c$ (dashed line) mark the phase transitions from NP to three superradiant phases. (c) Ground state energy $E_G(\alpha,\alpha^*)$ as a function of $\alpha_n$ for different quantum phases. (d) Energy spectrum $\varepsilon_{\pm}(k)$ for different coupling ratio $g_2/g_1$. (e)  Zak phase ($Z/\pi$) as a function of $J_2/J_1$ and $g_2/g_1$ with $g_1=1.5g_c$. $Z/\pi$ changes from $0$ to $1$ are marked by red (black) solid lines at $g_{\text{T2}}$ and $g_{\text{T3}}$. In all our calculations, we set $\omega=1$ as the units for
frequency, and $\Delta=50$, $J_1=0.1$. }
  \label{Fig1}
\end{figure}

\emph{Model and approach --}-We consider one-dimensional periodic array of coupled cavities with $N$ unit cells. Each unite cell contains two cavity sublattices labeled as $A_n$ and $B_n$. Each cavity are strongly coupled to a individual two-level atom of frequency $\Delta$ via a quantum Rabi Hamiltonian, realized by coupled each qubit to a resonator with
alternating a capactive or inductive couplings in Fig.~\ref{Fig1}(a). It gives rise to two distinct light-atom interactions~\cite{PRL2011,RevModPhys.93.025005}, described by complex coupling amplitudes with species-dependent modulus $g_{a(b)}$ and phase $\varphi_{a(b)}$. When such cavities are arranged in an array with photon hopping, the system can realize a photonic analog of the SSH configuration~\cite{PhysRevLett.42.1698,PhysRevB.22.2099}. The Hamiltonian of the quantum Rabi array is given by
\begin{equation} 
  \begin{aligned} \label{Ham1}
  H_{RL}=& \sum_{n=1}^{N} \frac{\Delta}{2} \sigma_{a,n}^{z}+\omega a_n^{\dagger}a_n+g_{a}\left(e^{-i\varphi_{a}}a_n^{\dagger}+a_n e^{i \varphi_{a}}\right) \sigma_{a,n}^{x}\\
  &+\frac{\Delta}{2} \sigma_{b,n}^{z}+\omega b_n^{\dagger}b_n+g_{b}\left(e^{-i\varphi_{b}}b_n^{\dagger}+b_n e^{i \varphi_{b}}\right) \sigma_{b,n}^{x}\\
  &+J_{1}a_{n}^{\dagger}b_{n}+J_{2}b_{n}^{\dagger} a_{n+1}+\text {H.c.},
\end{aligned}
\end{equation}
where $a_n^{\dagger} (b_n^{\dagger})$ is the bosonic annihilation operator for the cavity A(B) of frequency $\omega$, and $\sigma_{n}^{k}$ ($k = x,y,z$) is the Pauli matrices.  The intra-cell and inter-cell hopping strengths between cavities are denoted $J_1$ and $J_2$, respectively. Crucially, generating nontrivial topology requires species-dependent phases satisfying $\varphi_a -\varphi_b=\pm\pi/2$, as detailed in the Supplemental Material (SM)~\cite{supple}. Such phases can be realized by coupling a qubit coupled to an LC resonator via a capacitor or an inductor in each unit, yielding $\varphi_a=0$ and $\varphi_b = \pi/2$. This setup effectively couples atoms in sublattices A and B to two quadratures of a bosonic mode (e.g., the electric and magnetic components of an electromagnetic field), corresponding to electric and magnetic couplings. The dimensionless coupling strengths are defined as  $g_{1(2)} \equiv g_{a(b)} / \sqrt{\Delta \omega}$. 

In the infinite frequency limit $\Delta/\omega\rightarrow\infty$, the single-cavity quantum Rabi model exhibits a quantum phase transition from a normal to a superradiant phase~\cite{PhysRevA.87.013826,PhysRevLett.115.180404,PhysRevLett.119.220601,PhysRevA.101.033827}, spontaneously breaking $Z_2$ symmetry associated with the parity of the excitation number. On the other hand, the SSH lattice is known for its topological properties and possesses a global $U(1)$ symmetry. This has profound consequences for the topological phases of our system associated with the superradiant phase transition induced by light-atom interactions.

As the coupling strength increases, photon modes A and B become macroscopically excited. The bosonic operators are shifted by displacing the cavity field A and B, i.e., $D^\dagger[\pm\alpha_n]a_nD[\pm\alpha_n]=a_n\pm\alpha_n$ and $D^\dagger[\pm\beta_n]b_n D[\pm\beta_n]=b_n\pm\beta_n$, where the displacement operators are given by $D[\alpha_n]=e^{\alpha_n(a_n^\dagger-a_n)}$ and $D[\beta_n]=e^{\beta_n(b_n^\dagger-b_n)}$ with complex displacements $\alpha_{n}=A_{n}+iB_{n}$ and $\beta_n = X_n + iY_n$. The Hamiltonian $H_{RL}$ of Eq.(\ref{Ham1}) becomes (see SM~\cite{supple})
\begin{equation}
     \begin{aligned}\label{H_RL_SP}
         H_{RL}^{SP} = & \sum_{n=1}^{N}\omega a_{n}^{\dagger}a_{n} +g_a^{\prime}(a_{n}^{\dagger}+a_{n})\tau_{n,a}^x +\frac{\Delta_n^a}{2}\tau_{n,a}^z\\
         & \left.  + \omega b_{n}^{\dagger}b_{n}-ig_b^{\prime}(b_{n}^{\dagger}-b_{n})\tau_{n,b}^x+\frac{\Delta_n^b}{2}\tau_{n,b}^z \right. \\
         & +J_{1}a_n^{\dagger}b_n+J_{2}a_{n+1}^{\dagger}b_n + \text {H.c.},
     \end{aligned}
\end{equation}
where the effective coupling strengths in cavities A(B) is $g_{a(b)}^{\prime} = g_{a(b)}\Delta/\Delta^{a(b)}_{n}$. The transformed Pauli matrix are $\tau_{n,A}^z=\Delta/\Delta_n^{a}\sigma_n^z+4gA_n/\Delta_n^a\sigma_n^x$ with the renormalized atomic transition frequencies $\Delta^{a}_{n} = \sqrt{\Delta^2 + 16g_a^2A_n^2}$  and $\Delta^{b}_{n} = \sqrt{\Delta^2 + 16g_b^2Y_n^2}$, respectively. 

The ground-state energy is derived as $E_G=\sum_{n=1}\omega\left(|\alpha_n|^2 + |\beta_n|^2\right)-\Delta^{a}_{n}/2-\Delta^{b}_{n}/2 + 2J_{1}\left( A_nX_n + B_nY_n\right)+2J_{2}\left( A_{n+1}X_n + B_{n+1}Y_n\right)$. The displacement $\alpha_n$ and $\beta_n$ are determined by minimizing the ground-state energy, which also make the vanish of the off-diagonal terms in $H_{RL}^{SP} $ vanish. It yields $B_n=  -(1/\omega)(J_1+J_2)Y_n$ and $X_n=  -(1/\omega)(J_1+J_2)A_n$. Substituting in the ground-state energy giving $E_G(\alpha,\alpha^*)$ and minimizing it, the critical coupling strength is analytically obtained as $g_{1c}=g_{2c}=g_{c}=\sqrt{\omega^2 -(J_1+ J_2)^2}/(2\omega)$ (see SM~\cite{supple}). 

The phase diagram shown in Fig.\ref{Fig1}(b) features distinct superradiant phases characterized by the order parameters $\alpha_n$. (i)A normal phase (NP) exhibits $\alpha_n=\beta_n=0$ for $g_1<g_{c}$ and $g_2<g_{c}$. The energy has a minimum
in the origin. (ii) A superradiant "magnetic" phase (SMP) emerges for $g_1<g_c$ and $g_2>g_{c}$. The energy has two minima in the imaginary axis in Fig.\ref{Fig1}(c), giving a complex $\alpha_n$. It means that photons in sublattice A are displaced along the $y$ direction. (iii) A superradiant "electric" phase (SEP) is obtained for $g_2<g_c$ and $g_1>g_{c}$. The energy exhibits two minima in the real axis, giving a real $\alpha_n$. The bosons in sublattice A are displaced along the $x$ direction. (iv) A superradiant electromagnetic phase (SEMP) is obtained for $g_1>g_{c}$ and $g_2>g_{c}$. The ground state features complex $\alpha_n$, which corresponds to the displacement configuration in the $xy$ plane.

We perform a Schrieffer-Wolff transformation $U=\exp[-ig_{a}^{\prime}\tau_n^y/\Delta_{n}^{a}(a_{n}^{\dagger}+a_{n})]\exp[g_{b}^{\prime}\tau_n^y/\Delta_{n}^{b}(b_n-b_n^{\dagger})]$ to eliminate the $\tau_n^x$ terms in Hamiltonian (\ref{H_RL_SP}), which decouples the interactions between photon and atom. Projecting to the subspace of the lower atomic level $|\downarrow\rangle$, the low-energy effective Hamiltonian in momentum space is given by  (see SM~\cite{supple})
\begin{equation}
     \begin{aligned}\label{RL_SP}
         H_{RL}^{SP\downarrow} = & \sum_{k}\left[\omega_{a}^{\prime}a_{k}^{\dagger} a_{k} -\frac{g_a^{\prime2}}{\Delta^{a}_{n}}\left(a_{k} a_{-k}+a_{k}^{\dagger} a_{-k}^{\dagger}\right)\right. \\
          & \left. +\omega_{b}^{\prime}b_{k}^{\dagger} b_{k} + \frac{g_b^{\prime2}}{\Delta^{b}_{n}}\left(b_{k} b_{-k}+b_{k}^{\dagger} b_{-k}^{\dagger}\right)\right. \\
          & \left.+h(k) e^{-i\theta} a_{k}^{\dagger}b_{k}+h(k) e^{i\theta} b_{k}^{\dagger} a_{k} \right],
     \end{aligned}
\end{equation}
where $a_{n}^{\dagger}=\sum_{n} e^{i n k} a_{k}^{\dagger} / \sqrt{N}$ and $b_{n}^{\dagger}=\sum_{n} e^{i n k} b_{k}^{\dagger} / \sqrt{N}$ are the Fourier transformation. The renormalized cavity frequencies for cavity mode A and B are $\omega_{a}^{\prime} = \omega - 2g_{a}^{\prime2}/\Delta^{a}_{n}$ and $\omega_{b}^{\prime} = \omega - 2g_{b}^{\prime2}/\Delta^{b}_{n}$, and the effective hoping strengths is $h(k) e^{i\theta}=J_{1}+J_{2} e^{-i k}$. 
The Hamiltonian can be diagonalized by a Bogoliubov transformation 
$ H_{RL}^{SP\downarrow}(k) = \sum_k\varepsilon_{+}(k)\gamma^{+\dagger}_{k}\gamma^{+}_k + \varepsilon_{-}(k)\gamma^{-\dagger}_{k}\gamma^{-}_k$,
where the quasiparticle operators $\gamma_k^{\pm}$ is a linear combination of $\Phi_k = \left\{a^{\dagger}_k, b^{\dagger}_k, a_{-k}, b_{-k}\right\}$. The transformation $\Phi_k^{\dagger} = T\gamma_k^{\dagger}$ is implemented by a paraunitary matrix $T$, which satisfies the relations $T^{\dagger }\Lambda
T=\Lambda$ with $\Lambda =\sigma_z\otimes \mathbb{I}_{2}$ and $I_{2}$ being the identity matrix to preserve the bosonic commutation relations.

\begin{figure}
	\centering
	\includegraphics[width=\linewidth]{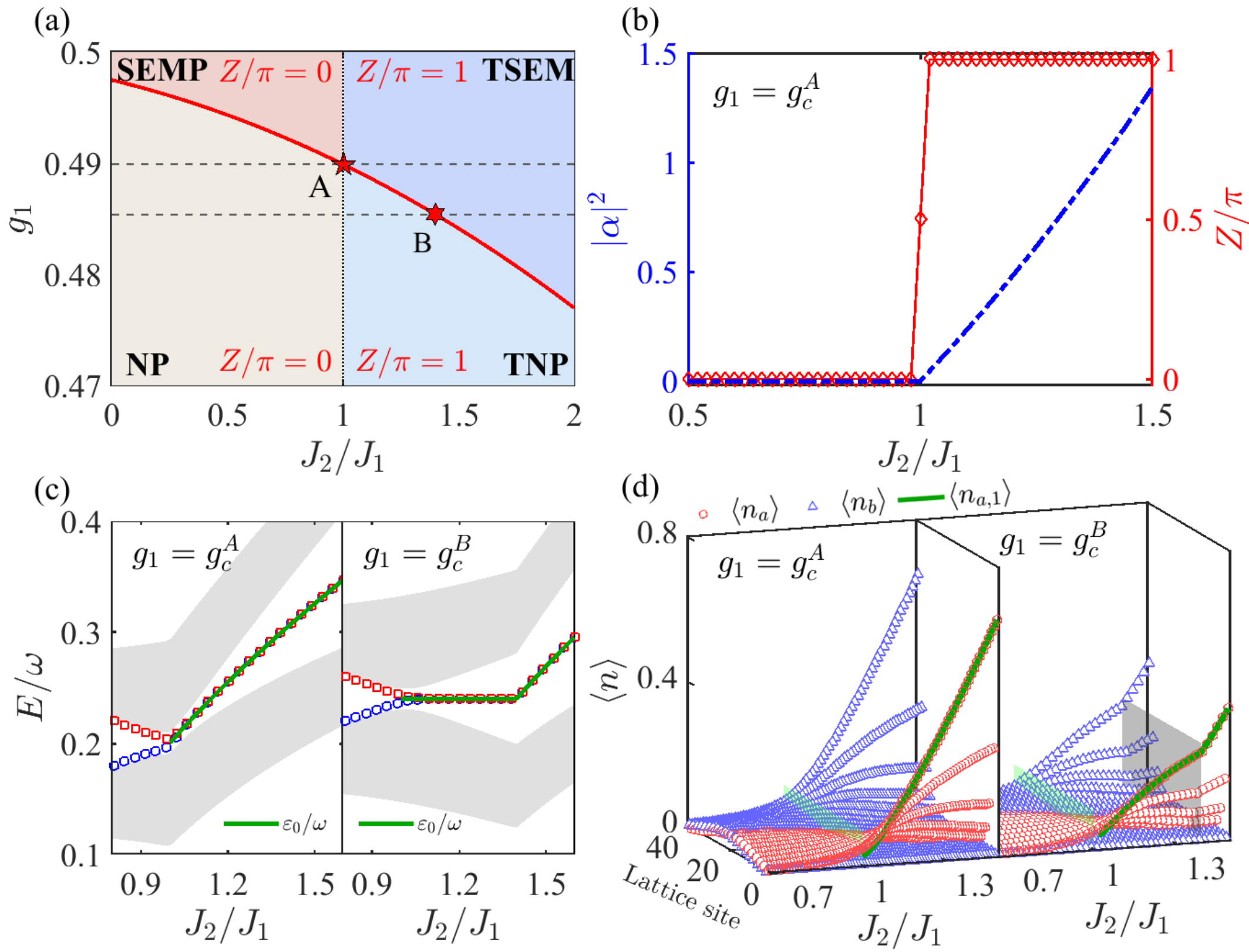}
	\caption{(a) Phase diagram in the $g_1$-$J_2/J_1$ plane for $g_1=g_2$. Nontrivial topological phases (TNP and TSEM; Zak phase $Z=\pi$) are separated from trivial ones (NP and SEMP; $Z=0$) by the critical boundary $g_{c}$ (red solid) and the topological transition line $J_2/J_1=1$ (black dashed). (b) Average photon number $|\alpha|^2$ of the sublattice A and Zak phase $Z/\pi$ as a function of $J_2/J_1$ for $g_1=g_c^{A}$ (point A).(c) Energy bands of $N=40$ unit cells at $g_1=g_c^{A}$ and $g_1=g_c^{B}$ (point B). Isolated eigenenergies agree well with the analytical result $\epsilon_0/\omega$ (green line). (d) Photon distributions at $g_1=g_c^{A}$ (NP-TSEM transition) and at $g_1=g_c^{B}$ (NP-TNP-TSEM transition). The photon number at site $1$ matches the analytical result $n_{a,1}$ (green line).}
	\label{Fig2}
\end{figure}

\emph{Topological Superadiant phases and Zak phase --}An outstanding feature of the system is the emergence of topology in the superradiant regime by tuning atom-cavity couplings. The nontrivial topological phase
emerges when the bulk energy gap closes. In general, the pairing particle-nonconserving terms in Hamiltonian (\ref{RL_SP}), e.g. $a^{\dagger}_ka^{\dagger}_{-k}$, $a_ka_{-k}$, do not close the energy gap~\cite{PhysRevB.97.041106}. This is the main difficulty that prevents the observation of superradiant phases with nontrivial topology.  Nevertheless, the closure of band gap can be tunable by the coupling strengths in a manner that differs from the $J_1/J_2$-dependent gap closing in the SSH model. 

The excitation spectrum $\varepsilon_{\pm}(k)$ of the Hamiltonian in Eq.(\ref{RL_SP}) are given explicitly as
\begin{equation}\label{Excitation_Sp_S}
		\varepsilon^2_{\pm}(k) =\left[\omega(\omega_a^{\prime}+\omega_b^{\prime})-\omega^2+h^2(k)\pm \delta \right],
	\end{equation}
	where $\delta = \omega\sqrt{(\omega'_b-\omega'_a)^2 +4h^2(k)(\omega_a^{\prime}+ \omega_b^{\prime}-\omega)/\omega}$.  
The vanishing of energy gap between $\varepsilon_{\pm}$ requires $\delta=0$, yielding $h(k)=0$ and $\omega_a^{\prime} =\omega_b^{\prime}$. We obtain the gap-closing at
$J_1 = J_2$ from $h(k)=0$ as in the SSH case. Additionally, our system requires identical on-site frequencies of two sublattices A and B, $\omega_a^{\prime} =\omega_b^{\prime}$, which preserves the inversion symmetry. It yields a topological critical coupling ratio $g_{\text{T1}}=g_2/g_1=1$ for SEMP and NP, which are marked in red circles in Fig.~\ref{Fig1}(b). For SMP, the gap closing occurs at (see SM~\cite{supple})
\begin{equation}\label{topoint2}
 g_{\text{T2}}=[1 - (J_1/\omega+J_2/\omega)^2]^{3/4}/\sqrt{8g_{1}^3}.  
\end{equation}
In contrast, SEP exhibits a distinct critical value 
\begin{equation}\label{topoint3}
g_{\text{T3}}=[1 -(J_1/\omega+J_2/\omega)^2]^{3/2}/8g_{1}^3. 
\end{equation}
These topological points are indicated by blue circles for $J_1/J_2=1$ in Fig.~\ref{Fig1}(b). The band structure of $\varepsilon_{\pm}(k)$ exhibits the gap closing for $k=\pi$ at the topological critical values, such as $g_{T1}$ and $g_{T3}$ in Fig.~\ref{Fig1}(d).

For isotropic coupling $g_a = g_b=g$, the Hamiltonian $ H_{RL}^{SP\downarrow}$ (\ref{RL_SP}) then maps onto the SSH model by make the pairing squeezing terms vanish. We perform the unitary transformation $S[\xi_{SR}^a]=\exp[\xi_{SR}^a(a^{\dagger}_ka^{\dagger}_{-k} - a_ka_{-k})/2]$ and $S[\xi_{SR}^b]=\exp[-\xi_{SR}^b(b^{\dagger}_kb^{\dagger}_{-k}-b_kb_{-k})/2]$ with the squeezing parameter $\xi_{SR}^a=\xi_{SR}^b$ (see SM). The transformed Hamiltonian then takes the SSH form (see SM)
\begin{equation}\label{RABISSH}
    \begin{aligned}
        H_{RS}= \varepsilon_{0}I_2 + (J_1 + J_2\cos{k})\sigma_x + J_2\sin{k}\sigma_y,
    \end{aligned}
\end{equation}
where the effective on-site energy is $\varepsilon_0= \omega_a^{\prime}\cosh{2\xi_{SR}^a}-2g_a^{\prime2}\sinh{2\xi_{SR}^a}/\Delta_n^a$. Unlike the SSH model, the on-site energy term $\epsilon_0$ breaks the chiral symmetry due to $PH(k)P^{\dagger}\neq-H(k)$ ($P=\sigma_z\otimes I_2$). It leads to non-symmetric spectrum with respect to $\varepsilon_0$. The corresponding Bogoliubov quasiparticle operators are analytically obtained by $\gamma_k^{\pm} = T_{1(2)}(k) \cdot (a_k, b_k, a_{-k}^{\dagger}, b_{-k}^{\dagger})$ with the corresponding eigenvector of the paraunitary matrix $T_{1(2)}(k) = \frac{1}{\sqrt{2}} \left(\pm e^{-i\theta} \cosh\xi_{SR}^a,  \cosh\xi_{SR}^b, \pm e^{-i\theta}\sinh\xi_{SR}^a, -\sinh\xi_{SR}^b\right)$.

The Zak phase can be interpreted as a topological invariant. The nontrivial topology of one-dimensional bipartite lattices is characterized by the Zak phase~\cite{PhysRevLett.62.2747}
\begin{equation}\label{Zak}
	Z^{\pm} =  i \int_{-\pi}^{\pi} T_{1(2)}^{\dagger}(k)*\partial_k T_{1(2)}(k) dk,
\end{equation}
which is determined by the distribution of photon momentum states $T_{1(2)}(k)$ in the Brillouin zone. Because the Bogoliubov transformation matrix $T$ is paraunitary, the inner product is modified such that $T_{1(2)}^{\dagger}(k)*\partial_k T_{1(2)}(k)=T_{1(2)}^{\dagger}(k)\Lambda\partial_k T_{1(2)}(k)$~\cite{PhysRevB.97.041106}. 
It is worth noting that for different on-site energies $\omega_a'$ and $\omega_b'$ breaks inversion symmetry, the Zak phase is not quantized and a topological number can not be defined. However, the inversion symmetry recovers when $\omega_a'=\omega_b'$ at the topological critical points $g_{T}$. The Zak phases is quantized, $Z^{\pm} = \int_{-\pi}^{\pi} \partial_k \theta(k)dk/2=\pm\pi$. Fig.~\ref{Fig1}(e) shows that Zak phase jumps from $0$ to $\pi$ at the topological critical point $g_{\text{T3}}$, signaling a nontrivial topological superradiant electric (TSE) phase. It also gives rise to a nontrivial topological superradiant electromagnet (TSEM) phase at $g_{\text{T1}}=1$. In contrast, when $g_2/g_1$ deviates from $g_{T}$, the Zak phase is not quantized due to the breaking of the inversion symmetry. 

For the SSH-type Hamiltonian $H_{\text{RS}}$, Zak phase distinguishes different regimes in Fig.~\ref{Fig2}(a): $Z=0$ marks the trivial SEMP and NP phases for $J_2/J_1<1$, while $Z=\pi$ identifies the nontrivial TSEM and topological norm phase (TNP) for $J_2/J_1>1$. A topological superradiant phase transition from NP to TSEM phase occurs at the critical coupling value $g_c$ and $J_2/J_1=1$ (point A) in Fig.~\ref{Fig2}(a). Especially, the transition is characterized by the monotonic increase of the intracavity photon number $|\alpha|^2$ in Fig.~\ref{Fig2}(b), accompanied by a quantized jump in the Zak phase from $0$ to $\pi$. Fig.~\ref{Fig2}(c) shows the nontrivial band topology in the TSEM phase for a finite system with $N=40$ unit cells. Two isolated eigenenergies in the midgap appear in the topological regime for $J_2/J_1>1$, which exhibit distinct dispersions versus $J_2/J_1$ in a sequence of the onsite energy $\varepsilon_0$ in Eq.(~\ref{RABISSH}). In contrast, for $g_1=g_c^B$ (point B), the system first undergoes the topological phase transition from NP to TNP, followed by a superradiant phase transition from TNP to TSEM phase.

\begin{figure}
  \centering
  \includegraphics[width=\linewidth]{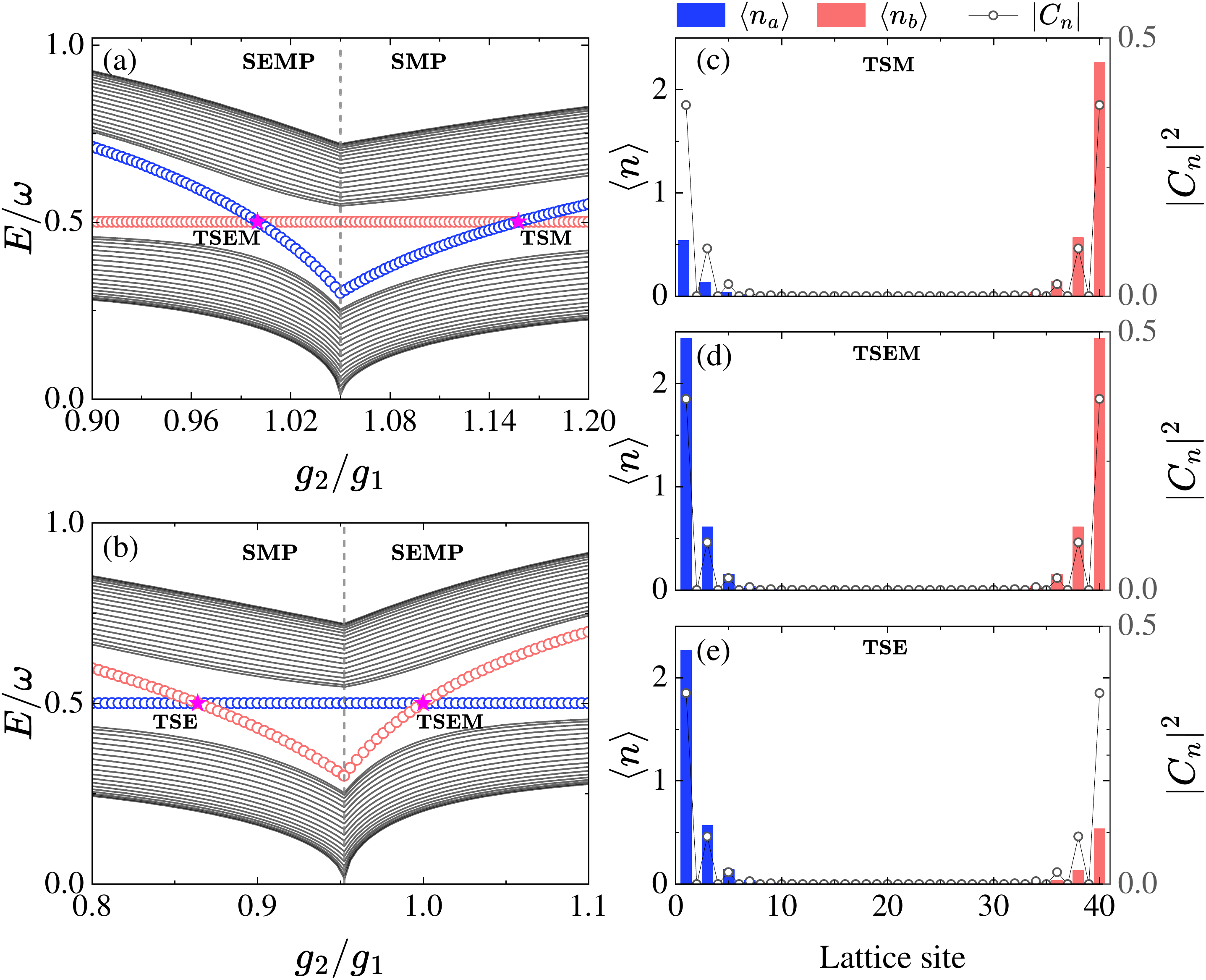}
  \caption{Energy spectrum of a finite-sized $N=40$ unit cells as a function of $g_2/g_1$ from SEMP to SMP phases with $g_2=1.05g_c$ (a), from SEP to SEMP phases (b) with $g_1=1.05g_c$ for $J_2/J_1>1$. Two isolated band-closing points identify the TSEM, TSM, and TSE phases. Average photon distributions $n_{a}$ and $n_b$ and real-space coefficients $|C_n|$ of edge modes in the TSM (c), TSEM (d) and TSE (e) phases. }
  \label{Fig3}
\end{figure}

\emph{Detecting edge states --}Bulk-edge correspondence guarantees the existence of a pair of midgap states at the boundaries of a finite topological system, which can be directly observed
in the photon excitations of the cavities. The edge operators of the effective Rabi-SSH Hamiltonian $H_{RS}$ in (\ref{RABISSH}) can be expressed in the SSH-like ansatz $\gamma_a = \sum_{n}c_n a_{n}$ and $\gamma_b= \sum_{n}d_nb_{n}$, where the coefficients are $c_n = \mathcal{N}\left(-J_1/J_2\right)^{n-1}/\sqrt{2}$ and $d_n =\mathcal{N}\left(-J_1/J_2\right)^{N-n}/\sqrt{2}$ with the normalization factor $\mathcal{N}$. $\gamma_{a/b}^{\dagger}$ creates edge-state excitations with energy $\varepsilon_{0}$, and satisfies the commutation relation $[H_{RS}, \gamma_{a/b}^{\dagger}] = \varepsilon_{0}\gamma_{a/b}^{\dagger}$. Considering the displacement and squeezing transformations on operators $a$ and $b$, the edge operators in the original basis can be expressed as 
$\gamma_a^\dagger=\sum_{n}c_n[\cosh \xi_{\text{SR}}^a(a_n^\dagger\pm\alpha_n^*)+\sinh \xi_{\text{SR}}^a(a_n\pm\alpha_n)]$ and $\gamma_b^\dagger=\sum_{n}d_n[\cosh \xi_{\text{SR}}^b (b_{n}^\dagger\pm\beta_n^*) - \sinh \xi_{\text{SR}}^b (b_{n}\pm\beta_n)]$(see SM) .
These can simplify to the edge operators in the TNP, where $\alpha_n=\beta_n=0$. 

For the TSEM phase, the edge state in the basis of the original Hamiltonian is obtained as (see SM~\cite{supple})
\begin{equation}
 |\Psi_{\text{TSEM}}^{\pm}\rangle=[\pm \gamma_a^\dagger+\gamma_b^\dagger]\tilde{|\mathrm{G}\rangle}_{\text{TSEM}}|\tilde{\downarrow}_a\rangle\otimes|\tilde{\downarrow}_b\rangle,   
\end{equation}
where $\tilde{|\mathrm{G}\rangle}_{TSEM}=D[\pm\alpha_n]D[\pm\beta_n]S[\xi_{SR}^a]S[\xi_{SR}^b]|\mathrm{G}\rangle$ represents a displaced-squeezed state. And $|\tilde{\downarrow}_a\rangle$ ($|\tilde{\downarrow}_b\rangle$) is the eigenstate of the transformed Pauli matrix $\tau_{n,a(b)}^z$ in Eq.(\ref{H_RL_SP}). 

The average photon number of edge modes in sublattices A and B on site $n$ are derived as
\begin{eqnarray}\label{analytical}
		n_{a,n} &=&|c_n|^2/2\{[3\cosh(2\xi_{SR}^a)-1]/2 + |\alpha_n|^2\},\\
		n_{b,n} &=&|d_n|^2/2\{[3\cosh(2\xi_{SR}^b)-1]/2 + |\beta_n|^2\}.
\end{eqnarray}
The contributions from $\alpha_n$ and $\beta_n$ terms has profound consequence for the excitation of edge modes in the superradiant regime.

Fig.~\ref{Fig2} (d) shows the distribution of average photons at the critical value $g_c^{A}$ of point A for $g_1=g_2$. For $J_2/J_1<1$, photon excitations reside in bulk eigenstates and exhibit a spatial profile that decays exponentially toward the boundaries. When $J_2/J_1>1$, the mean  photon excitations of edge modes $n_{a,n}$ ($n_{b,n}$) are localized near the boundaries of sublattices $(1,A(B))$ and $(40,A(B))$, and grow monotonically, which signals the topological superradiant phase transition at $g_c^{A}$. In contrast, at $g_{c}^{B}$ of point B, photon excitations of two edge modes appear in the TNP regime in Fig.~\ref{Fig2}(d), then rises sharply in the TSEM regime. Consequently, edge-mode excitations in the TSEM phase show a pronounced nonlinear enhancement by comparing to the TNP. This amplified excitations indicate superradiance-enhanced edge-mode excitations.

For the topological superradiant magnetic (TSM) and 
TSE phases, we perform an exact Bogoliubov diagonalization in real space for a finite-sized $N=40$ unit cells. The coupling ratio $g_2/g_1$ closes the superradiance-induced band gap at two critical points $g_{T1}$ and $g_{T2}$ ($g_{T3}$) while remaining open elsewhere, which identify the TSEM, TSM and TSE phases in Fig.~\ref{Fig3}(a)-(b). They have the same coefficients distribution in contrast to photons distribution in Fig.~\ref{Fig3}(c)-(e). In the TSEM phase, photon distributions localized at either the left or right boundary, decaying exponentially into the bulk. By contrast, in the TSM phase, the photon distribution localizes exclusively on the right edge, decaying exponentially on sublattice B near sites ($40,B$) in Fig.~\ref{Fig3}(c). Conversely, in the TSE phase, the photon excitations localize at the left boundary on sublattice A near sites ($1,A$) in Fig.~\ref{Fig3}(e). Thus, the edge modes in the TSE and TSM phases lead to robust unidirectional superradiant-enhanced excitation of photons, exhibiting opposite chirality.

\emph{Conclusion --}To summarize, we uncover anomalous topological superradiant phases of the quantum Rabi array. The complex light-atom coupling restores inversion symmetry to inherit topology of the SSH lattice, and close superradiance-induced band gap in a novel manner. In topological superradiant electric (magnetic) phase, the chiral edge states exhibit superradiance-enhanced amplification of photon excitations localized in the left or right boundary. The marriage between quantum optics and topology promises new advance in quantum technologies, such superradiant lasing~\cite{sbbk-xdvs}.  

\textit{Acknowledgments--}
This work was supported by National Natural Science Foundation of China Grant No.12475013, No. 12547101, and the Fundamental ResearchFunds for the Central Universities Grant No. 2025CDJ-IAISYB-029.
 

\bibliography{apssamp}

\end{document}